\documentclass[12pt]{article}
\usepackage[utf8]{inputenc}
\usepackage{setspace}
\usepackage{xspace}
\usepackage{graphicx}
\usepackage{bm}
\usepackage{courier}
\usepackage{appendix}
\usepackage{chngcntr}
\usepackage{amsmath}
\usepackage{booktabs}
\usepackage{verbatim}
\usepackage{multirow}
\usepackage{multicol}
\usepackage{subfigure}
\usepackage{caption}
\usepackage{float}
\usepackage{optidef}
\graphicspath{{images/}}

\newcommand{\startparent}{%
  \setcounter{equation}{0}%
  \xdef\theparentequation{\arabic{parentequation}}%
}
\newcommand\wordcount{
    \immediate\write18{texcount -sub=section \jobname.tex  | grep "Section" | sed -e 's/+.*//' | sed -n \thesection p > 'count.txt'}
(\input{count.txt}words)}
\usepackage[a4paper,
            bindingoffset=0.2in,
            left=1in,
            right=1in,
            top=1in,
            bottom=1in,
            footskip=.25in]{geometry}
\usepackage{authblk}
\usepackage{appendix}
\usepackage[colorlinks=true,urlcolor=black,linkcolor=blue,citecolor=blue]{hyperref}
\usepackage{cite}
\renewcommand{\arraystretch}{1.4}
\usepackage[left]{lineno}

\title{Multi-trip algorithm for multi-depot rural postman problem with rechargeable vehicles}
\date{}

\author[a]{Eashwar Sathyamurthy}
\author[b]{Jeffrey W. Herrmann}
\author[a]{Shapour Azarm}
\affil[a]{Department of Mechanical Engineering, University of Maryland, USA}
\affil[b]{Institute for Systems Research, Department of Mechanical Engineering, University of Maryland, USA}

\begin{document}
\maketitle
\begin{abstract}
This paper studies an extension of the rural postman problem with multiple depots and rechargeable and reusable vehicles capable of multiple trips with capacity constraints. This paper presents a new Mixed Integer Linear Programming (MILP) formulation to find optimal solutions to the problem. The paper also proposes a new heuristic called the multi-trip algorithm for the problem whose solutions are compared against solutions of heuristics from literature and the optimal solutions obtained from the MILP formulation by testing them on both benchmark instances and real-world instances generated from road maps. Results show that the proposed heuristic was able to solve all the instances and produce better solutions than heuristics from the literature on 37 of 39 total instances. Due to the high requirement of memory and compute power, the Gurobi optimizer used for solving the MILP formulation, although it produced optimal solutions, was only able to solve benchmark instances but not real-world instances.
\end{abstract}
\hspace{0.9cm}\textbf{\textit{keywords-}} Heuristic, Multi-trip algorithm, Rural postman problem, Capacity constraints, Rechargeable and reusable vehicles.


\section{Introduction}
\label{intro}
This study of the multi-depot rural postman problem with rechargeable vehicles and reusable vehicles (MD-RPP-RV) was motivated by a real-world problem of planning routes for unmanned aerial vehicles (UAVs). In this problem, the UAVs were needed to travel from bases located at different locations in an area of interest (AOI) to road segments that might be icy. The information about possibly icy road segments is obtained from analyzing historical data on road conditions in the AOI and through expert opinion. But, the road segments need to be visually inspected the UAVs. \par
The vehicles' mission is to use their sensors to collect data about the possibly icy roads and confirm that they are icy, which can be used to warn drivers of hazardous conditions.
The vehicles must follow the roads as they visit the icy roads and return to their bases.
Moreover, the vehicles have a limited range (capacity) due to the limited energy in their batteries and hence must visit a base to recharge before the battery is depleted.
The road map in the AOI is represented as a graph in which bases are a subset of nodes called depots, and the possibly icy roads are a subset of edges called required edges that must be traversed. 

This paper formulates the MD-RPP-RV as a mixed integer linear progam (MILP), presents a heuristic-based algorithm to find feasible solutions, and describes the results of a study that compares its performance with the optimal solutions obtained from solving the MILP formulation using the Gurobi optimizer \cite{gurobi}. \par

The MD-RPP-RV is related to variants of the Capacitated Arc Routing Problem (CARP) \cite{Carp}.
The Chinese Postman Problem (CPP) and Rural Postman Problem (RPP) are well-known variants of the CARP, both aimed at determining the shortest cycle tour of minimum length for a single postman. CPP \cite{cpp} requires that all edges $E$ must be traversed at least once, and RPP \cite{rpp} requires a subset of $E$ to be traversed. Despite many similarities between CPP and RPP, they have different computational complexities. CPP can be solved optimally in polynomial time using a matching algorithm \cite{complexcpp}, but the RPP is NP-hard\cite{rppnphard}. 
MD-RPP-RV is NP-hard to solve because it can be reduced to the RPP by considering only one depot and restricting vehicles to a single trip. \par
 
The MD-RPP-RV is a combination of multi-trip \cite{Carp14,Carp15} and multi-depot \cite{mdcarp, mdcarp1, mdcarp2} variants  of the CARP. In the multi-trip CARP, each vehicle is capable of performing multiple trips to traverse all required edges.
In the multi-depot CARP, vehicles can start from multiple depots. 
MD-RPP-RV most resembles the multi-depot RPP (MD-RPP), in which vehicles with limited capacity located at multiple depots must traverse a subset of edges.
In the MD-RPP \cite{constraint12}, the vehicles start and end at the same depot and are restricted to a single trip, which is not the case in the MD-RPP-RV.
The MD-RPP-RV can represent many real-world problems, such as routing vehicles optimally for snowplowing or routing UAVs to inspect leaky gas lines or deformed railway tracks. \par

In the literature, three common approaches exist to solve arc routing problems. These approaches include heuristic, metaheuristic, and exact approaches. Heuristic-based approaches usually generate a sub-optimal solution without a guarantee of its quality. Frederickson \textit{et al.} \cite{mmcpp} proposed the first constructive heuristic for the min-max CPP. 
Ahr \textit{et al.} \cite{mmcpp1} proposed the Augment-Merge and Cluster heuristics, which found better solutions in most instances.
Although the solutions might be sub-optimal, these heuristics can quickly produce feasible solutions, which is useful when dealing with large instances (those with many vehicles and arcs).

Metaheuristic approaches \cite{Carp7, Carp9, Carp13, Carp14, Carp17, Carp18} are generalized methods that can be used to solve a wide range of different problems by tuning their hyperparameters to make them suitable to solve a specific problem. 
Typically, these approaches start with an initial solution generated from the heuristic-based approach, explore the solution space, and converge to a new solution. 
Hertz \textit{et al.} \cite{Carp6} proposed a tabu search heuristic called CARPET, which produced better solutions than heuristic-based approaches \cite{Carp2} and optimal solutions in most benchmark instances. Lacomme \textit{et al.} \cite{Carp12} proposed the first genetic algorithm (GA) for the CARP, and the results showed that the GA solutions were better than CARPET solutions on benchmark instances with reduced computational time. 
Tuning the hyperparameters can be a lengthy, trial-and-error process, however, these methods need an initial solution, and their computational time can vary depending on the quality of the initial solutions and stochasticity of the solution process.\par

Exact approaches \cite{Carp20, Carp23, Carp25, Carp26} produce optimal solutions by formulating the problem as a MILP and solving it using branch-and-bound techniques \cite{brandbo}. 
Fernández \textit{et al.} \cite{mdrpp} studied the MD-RPP on undirected graphs and presented MILP formulations with two different objective functions: (i) minimize total routing costs and (ii) a min-max objective function aiming at minimizing the length of the longest route.
They proposed a branch-and-cut algorithm that produced balanced routes for both objectives. 
Because finding and verifying optimal solutions requires examining (at least implicitly) all feasible solutions, these approaches can become impractical to use for large instances due to their computation and memory requirements.

The primary need for developing the multi-trip algorithm can be attributed to the popularity of autonomous vehicles and their ability to perform surveillance operations \cite{sur1, sur2, sur3}. 
Routing vehicles require planning routes considering their battery limitations and frequent recharging and reusing to accomplish the task effectively. 
The proposed multi-trip algorithm addresses these limitations by integrating the battery and charging constraints. 

The main contributions of this paper are the following:
\begin{enumerate}
    \item The paper introduces a new class of arc routing problem called MD-RPP-RV in which the vehicles are rechargeable and reused to perform multiple trips from multiple depots that can end up at different depots. Although using rechargeable and reusable vehicles capable of multiple trips is explored in the literature from the perspective of node routing problems \cite{sur4, sur5, sur6, sur7, sur8}, and reusable vehicles in arc routing problems like multi-trip CARP \cite{Carp14, Carp15}, the idea of integrating rechargeability and reusability aspects of vehicles to solve multi-depot arc routing problems using multiple trips has not been looked at in the literature to the best of our knowledge. 
    
     \item The paper formulates the problem as a MILP formulation, which can be used to find optimal routes. 

    \item The proposed multi-trip algorithm can quickly solve instances, including those based on real-world roadmaps (Figure \ref{rw-results}). Results from testing the multi-trip algorithm, existing CARP heuristics \cite{Carp2, Carp3, Carp5}, and the Gurobi optimizer show that the multi-trip algorithm found better solutions than the other  CARP heuristics for 37 of 39 instances.

   
\end{enumerate}

The remainder of this paper is organized as follows:
Section \ref{profor} provides the assumptions considered and presents a MILP formulation for the MD-RPP-RV. Section \ref{sa} briefly describes the proposed multi-trip algorithm. Section \ref{rd} provides the results obtained from testing the multi-trip algorithm and other CARP algorithms on instances for the MD-RPP-RV. Section \ref{conc} concludes the paper by providing a summary.


\section{Problem Formulation and Assumptions}
\label{profor}
This section describes the assumptions considered to help convert the problem of routing vehicles to collect road information into a combinatorial optimization problem.

\subsection{Assumptions}
\label{assump}
For the MD-RPP-RV, we made the following assumptions:
\begin{enumerate}
    \item All vehicles start from their respective depots with full battery capacity.
    \item The depots at which vehicles start, stop, and recharge are located at intersections of roads.
    \item If an edge is required, at least one vehicle must traverse the edge. 
    \item All vehicles have the same capacity (operating time when fully charged).
    \item All vehicles move at the same speed.
    \item The time to recharge the vehicle battery is the same at all depots.   Partial recharging is not possible.
    \item The graph that represents the AOI is connected.
    \item The time required to traverse an edge depends only upon the length of the edge and the vehicle speed.
    \label{a6}
\end{enumerate}

The following problem formulation incorporates these assumptions.

\subsection{Problem Formulation}
\label{formulation}
Let $G = (N, E, T)$ be the undirected graph that models the AOI where $N$ is the set of nodes (locations), $E$ is the set of edges ($(i,j) \in E$) that connect nodes ($i,j \in N$), and $T$ is the set of edge weights ($t(i,j) \in T$). 
An edge represents a road segment; let $l(i,j)$ be the length of the road segment in meters. 
Let $S$ be the constant speed in meters per second at which all vehicles traverse road segments. Let $t(i,j)$, the edge weight, be the time taken by a vehicle that traverses edge $(i,j)$:  $t(i,j) = l(i,j) / S$. 

Let $E_u \subseteq E$ be the set of required edges that needs to be visited. Let $N_d$ be the set of depots where vehicles can start or stop or re-fuel (recharge).  $N_d \subseteq N$. 
Let $T_{max}$ be the maximum time that a vehicle remains operational after charging.
Let $K$ be the number of vehicles. 
Let $R_T$ be the time taken to recharge a vehicle. 
Let $F$ be the maximum number of trips that a vehicle can make determined using an iterative procedure (Section \ref{deter-trips}). 

The MD-RPP-RV can be formulated as the following MILP:
 
\newpage
\begin{equation*}
    \text{minimize} \quad \beta
\end{equation*}
\begin{alignat}{3}
\startparent
&\text{such that}
\sum_{(B(k),j) \in E} x(k, 1, B(k), j) = z(k, 1), \quad  k = 1,...,K\\
&z(k, f) - z(k, f+1) \geq 0, \quad k = 1,...,K, f = 1,...,F-1\\
&\sum_{(i,d) \in E, d \in N_d} x(k, f, i, d) = y(k, f, d), \quad k = 1,...,K,  f = 1,...,F\\
&y(k, f-1, d) \geq \sum_{(d,j) \in E} x(k, f, d, j), \quad k = 1,...,K,  f = 2,...,F, d \in N_d\\
&z(k,f) - \sum_{d \in N_d} y(k,f,d) = 0, \quad k = 1,...,K,  f = 1,...,F\\
&\sum_{f=1}^{F} \sum_{(i.j) \in E} x(k, f, i, j) \times t(i,j) + (\sum_{f=1}^{F} z(k,f) - 1) \times R_T \leq \beta, \quad k = 1,...,K\\
&\sum_{(i,j) \in E} x(k, f, i, j) \times t(i,j) \leq T_{max}, \quad k = 1,...,K,  f = 1,...,F\\
&\sum_{(i,j) \in E, i \in N_d} x(k, f, i, j) - \sum_{(i,j) \in E, j \in N_d} x(k, f, i, j) = 0,  \quad k = 1,...,K,  f = 1,...,F\\
&\sum_{j \in N} x(k, f, i, j) - \sum_{j \in N} x(k, f, j, i) = 0, \quad k = 1,...,K,  f = 1,...,F, i \in N/\{N_d\}\\
&\sum_{k=1}^{K} \sum_{f=1}^{F} x(k,f,i,j) + \sum_{k=1}^{K} \sum_{f=1}^{F} x(k,f,j,i) \geq 1, \quad \forall (i,j),(j,i) \in E_u\\
&\sum_{(i,j) \in E} x(k,f,i,j) \leq z(k,f) \times M, \quad k = 1,...,K,  f = 1,...,F \\
&\sum_{(i,j) \in \delta(S)} x(k,f,i,j) \geq 2 \times x(k,f,p,q), \; k = 1,...,K,  f = 1,...,F, \forall S \subseteq N/\{N_d\}, (p,q) \in E(S)\\
&x(k,f,i,j) \in [0,1],\quad k = 1,...,K,  f = 1,...,F, \forall (i,j) \in E\\
&y(k,f,d) \in [0,1], \quad k = 1,...,K,  f = 1,...,F, d \in N_d\\
&z(k,f) \in [0,1], \quad k = 1,...,K,  f = 1,...,F\\
&\beta \in R^{+}, \: M >> |E|
\end{alignat}
 
There are three sets of binary decision variables in the formulation. 
$x(k,f,i,j) = 1$ if vehicle $k$ traverses edge $(i,j)$ from node $i$ to node $j$ during its $f$-th trip.
$x(k,f,j,i) = 1$ if vehicle $k$ traverses edge $(i,j)$ from node $j$ to node $i$ during its $f$-th trip.
Note that a vehicle begins a trip at a depot, traverses a path through the graph, and ends the trip at a depot in order to recharge.
$y(k,f,d)$ keeps track of the depot nodes where each vehicle has ended its trips. It is 1 if vehicle $k$ ends its $f$-th trip at depot node $d \in N_d$, else, it is 0. 
$z(k,f)$ is 1 if vehicle $k$ vehicle uses its $f$-th trip; else it is 0. 
For $k = 1, \ldots, K$, $B(k)$ is a node in $N_d$ such that vehicle $k$ is initially at depot $B(k)$.  The values of $B(k)$ are given.  \par

The objective function $\beta$ is the maximum total time that any vehicle needs to complete all of its trips and the recharging between trips.
Constraint (1) ensures that if vehicle $k$ has initiated its first trip from depot node $B(k)$, it is tracked using $z(k,1)$. Constraint (2) restricts unnecessary trips by ensuring that trip $f+1$ is used only if trip $f$ is used by vehicle $k$. Constraint (3) keeps track of the depot node $d \in N_d$ where each trip $f$ of vehicle $k$ has ended using $y(k,f,d)$. Constraint (4) forces vehicle $k$ to start its flight $f$ from the same depot node $d \in N_d$ it ended its previous trip $f-1$ if necessary. Constraint (5) ensures that vehicle $k$, if it begins trip $f$, must end in any of the depot nodes $d \in N_d$. \par
Constraint (6) enforces an upper bound for maximum trip time for all vehicles considering recharge time. Note constraint (6) only adds recharge time for used trip $f > 1$ determined using the $z(k,f)$. Constraint (7) ensures that all trips taken by each vehicle $k$ satisfy the battery constraint. Constraint (8) ensures that all trips made by each vehicle enter or leave depot nodes the same number of times. Constraint (9) make sure all trips of each vehicle have a continuous flow. Constraint (10) ensures that all trips made by all vehicles traverse each required edge at least once.
Constraint (11) ensures that a vehicle traverses no edges during a trip that is not used.
Constraint (12) is a subtour elimination constraint that is based on one used by Chen \emph{et al} \cite{constraint12}.
The set $S$ is a set of nodes that are not depots, $E(S)$ is the set of required edges between nodes in $S$, and $\delta(S)$ is the set of edges that have only one node in the set $S$.
This constraint ensures that, if a vehicle traverses a required edge $(p,q) \in E(S)$, then it also traverses edges that enter and leave $S$. \par

The size of this MILP grows exponentially with the number of nodes, unfortunately.
For constraint (12) alone, there are $2^{|N/N_d|}$ subsets $S$, so the formulation requires a total of $2^{|N/N_d|} \times K \times F$ copies of this constraint.
The following section describes our multi-trip algorithm, which addresses the computational complexity problem and makes it  suitable for finding solutions for large instances.


\section{Solution Approach}
\label{sa}
This section first describes the multi-trip routes used by the algorithm to traverse required edges with the help of an example. Then, the pseudocode of the algorithm is explained step by step to help understand the working of the algorithm. 

\subsection{Multi-trip Routes}
\label{routetypes}
A feasible solution specifies a route for each vehicle.
A route includes one or more trips; each trip begins and ends at the same or different depot.
The length of a trip is limited by the capacity constraint.
Although its mission is to cover required edges, a vehicle might need to make a trip that includes no  required edges in order to get closer to one or more required edges.

\begin{figure}[H]
    \centering
    \includegraphics[width=15.5cm]{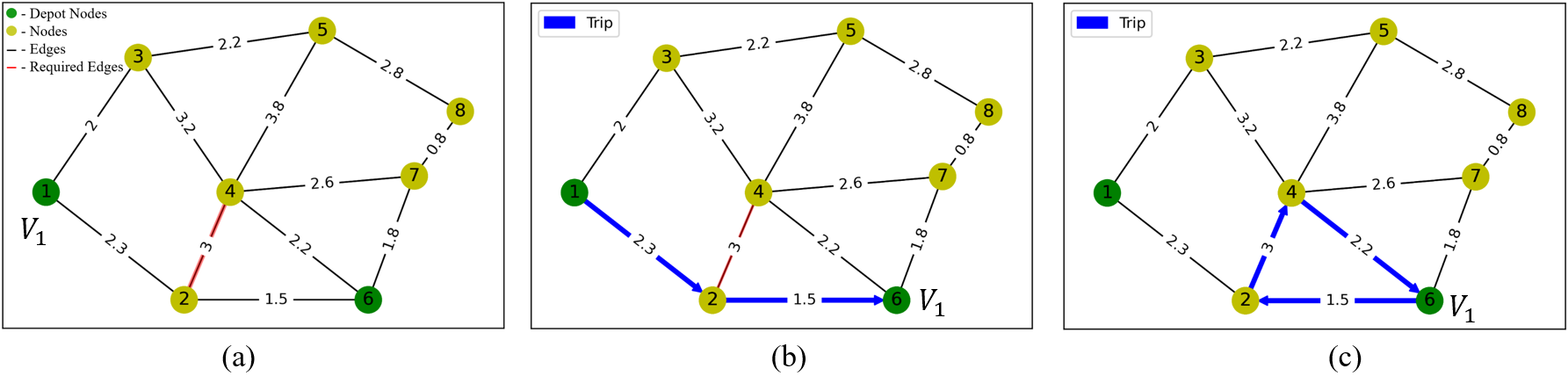}
    \caption{\textrm{ (a) A sample instance of MD-RPP-RV, (b) Vehicle $V_1$ getting closer to the required edge. (c) Vehicle $V_1$ traversing the required edge.}
    \label{case2}}
\end{figure}

Consider an instance of the problem consisting of an undirected graph $G$ with 8 nodes, 12 edges, 2 depot nodes, and 1 required edge, shown in Figure \ref{case2}{\color{blue}(a)}. 
There is only one vehicle ($V_1$); it has a capacity of 7 time units, it begins at depot node 1, and its recharging time equals 1.1 time units.

Due to the capacity constraint, the vehicle cannot directly traverse the required edge $(2, 4)$ in one trip. 
The best possible trip to cover this edge is \{1-2-4-6\}, but this requires 7.5 time units, which exceeds the vehicle capacity of 7 time units, so it is infeasible. 

 In a feasible multi-trip route, the vehicle travels from its current depot to another depot (node 6) to move closer to the required edge (Figure \ref{case2}{\color{blue}(b)}), recharges at that depot, and then completes a trip that includes the required edge (Figure \ref{case2}{\color{blue}(c})). 
 
The first trip \{1-2-6\} takes 3.8 time units for the vehicle to complete. Then, the vehicle recharges at node 6, which takes 1.1 time units. 
Finally, the vehicle takes the second trip \{6-2-4-6\}, which takes 6.7 time units to complete. 
Each trip is feasible because its duration is less than the capacity of 7 time units.
The vehicle requires 11.6 time units to complete the route and traverse the required edge, which is the sum of both trip times and the recharge time. 

\subsection{Multi-trip Algorithm}
\label{v1}

Table \ref{tb1} provides the nomenclature used in the pseudocode of the proposed algorithm.

\begin{table}[H]
\caption{\textrm{Nomenclature}}
\centering
\label{tb1}
\renewcommand{\arraystretch}{1}
\resizebox{\textwidth}{!}{\begin{tabular}{ccl}
\hline
Symbols                  & Type   & \multicolumn{1}{c}{Interpretation}                              \\ \hline
$G = (N, E, T)$  & Input  & Graph consisting of nodes,   edges, edge weights          \\ 
$N_d$             & Input  & Set of depots $ N \subseteq N_d$                                            \\ 
$E_u$              & Input  & Set of required edges                     \\ 
$K$               & Input  & Number of vehicles                                            \\ 
$t_k$, $k = 1, \ldots, K$ & Input  & Time at which vehicle $k$ is available                          \\ 
$P_k$, $k = 1, \ldots, K$ & Input  & Path for vehicle $k$ until time $t_k$                              \\ 
$n_k$, $k = 1, \ldots, K$ & Input  & Location of vehicle $k$ at time   $t_k$                           \\ 
$u_k$, $k = 1, \ldots, K$ & Input  & Utilization of vehicle $k$ (elapsed   time since last recharge) \\ 
$y_k$, $k = 1, \ldots, K$ & Input  & Time that vehicle $k$ arrived at last   depot in $P_k$             \\ 
$R_T$              & Input  & Time to recharge a vehicle at   a depot                           \\ 
$C$               & Input  & vehicle capacity (time)                                       \\ 
$P_k$, $k = 1, \ldots, K$ & Output & Path for vehicle $k$                                            \\ 
$y_k$, $k = 1, \ldots, K$ & Output & Time that vehicle $k$ arrived at last   depot in $P_k$             \\ \hline
\end{tabular}}
\end{table}

The multi-trip algorithm considers the location ($n_k$), utilization ($u_k$), and availability ($t_k$) of the vehicles to determine feasible, high-quality trips that traverse the required edges.


\begin{figure}[H]
    \centering
    \includegraphics[width=15cm]{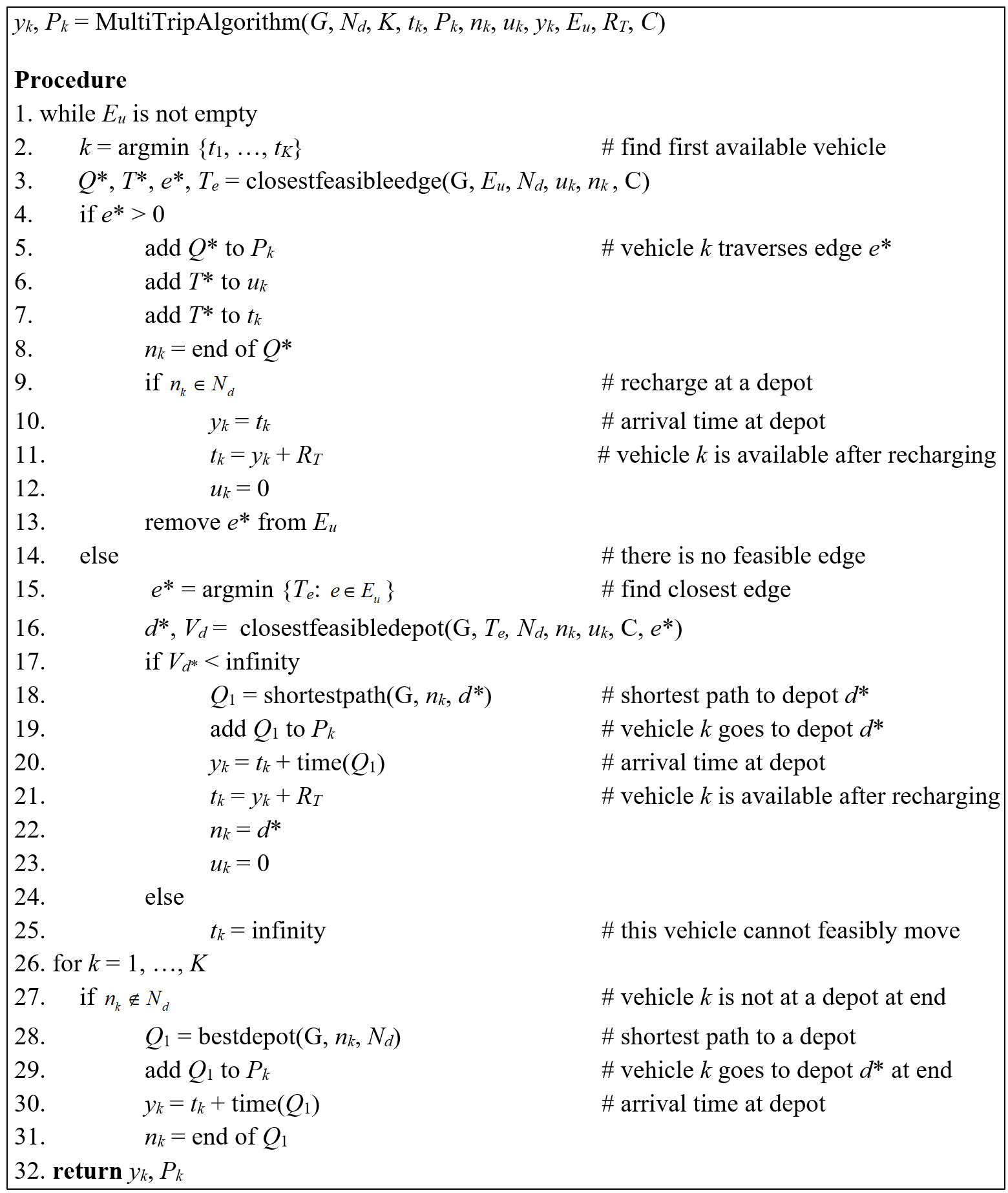}
    \caption{\textrm{Pseudocode for the Multi-trip Algorithm}}
    \label{sol2122}
\end{figure}

The multi-trip algorithm, Figure \ref{sol2122}, iterates until all of the required edges have been traversed (line 1). 
In each iteration, the algorithm determines which vehicle is available next (line 2) and the closest edge that the vehicle can reach in a single trip.
The algorithm then determines a path to traverse that edge using the $closestfeasibleedge()$ subroutine, which uses  Dijkstra's algorithm \cite{dijkstra} and updates the location ($n_k$),  utilization ($u_k$), availability ($t_k$) of the vehicle and the required edges ($E_u$) (lines 4-13). 


If a multi-trip route is necessary (lines 15-23), the algorithm first determines the required edge ($e^{*}$) that is closest to the vehicle location and moves the vehicle to a depot that is closer to $e^{*}$ using the $closestfeasibledepot()$ subroutine. 
The algorithm handles single-trip and multi-trip scenarios differently: in a single-trip scenario, the first available vehicle is allocated to a required edge; but in a multi-trip scenario, the closest required edge is allocated to the vehicle.

The algorithm makes a vehicle infeasible to move (line 17) if it cannot traverse a required edge or move to another depot (due to the capacity constraint).
(This might  happen when $G$ is disconnected.) 
After all of the required edges in $E_u$ are traversed, the algorithm completes the partial vehicle paths that do not end at depot nodes (lines 26-31).
Finally, the algorithm returns for each vehicle $k$ both $P_k$, the vehicle's path, and $y_k$, the time at which the vehicle arrives at the last depot in $P_k$. The following subsection demonstrates the working of the multi-trip algorithm with the help of an example.

\subsection{Demonstration Example}
\label{demo}
\begin{figure}[H]
    \centering
    \includegraphics[width=8.5cm]{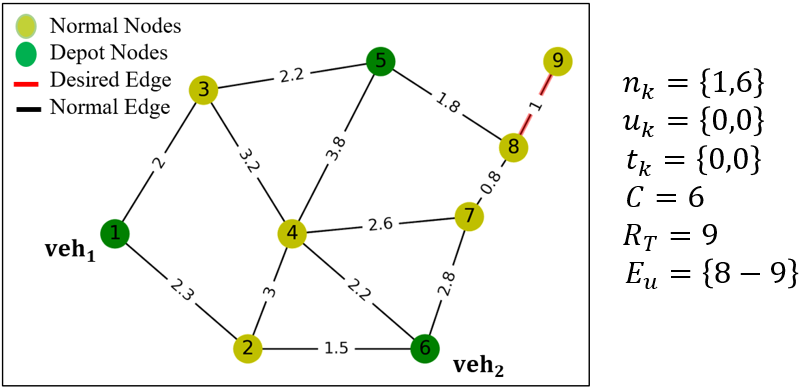}
    \caption{\textrm{Sample instance ($n_k = $ Vehicle Location, $u_k = $ Vehicle utilization, $t_k = $ Vehicle availability, $C = $ Vehicle capacity, $R_T = $ Recharge Time, $E_u$ = Required Edges)}}
    \label{sameexam1}
\end{figure}

Consider an instance with an undirected graph that has 9 nodes and 13 edges. 
As shown in Figure \ref{sameexam1}, this instance has one required edge $(8, 9)$, three depots (nodes 1, 5, and 6), and two vehicles, which begin at nodes 1 and 6.
The vehicle capacity is 6 time units, and the recharge time equals 9 time units. 
In Figure \ref{sameexam1}, the vehicle location ($n_k$), vehicle utilization ($u_k$) and vehicle availability ($t_k$) have two elements (\{veh$_1$, veh$_2$\}).

\begin{figure}[H]
    \centering
    \includegraphics[width=10.5cm]{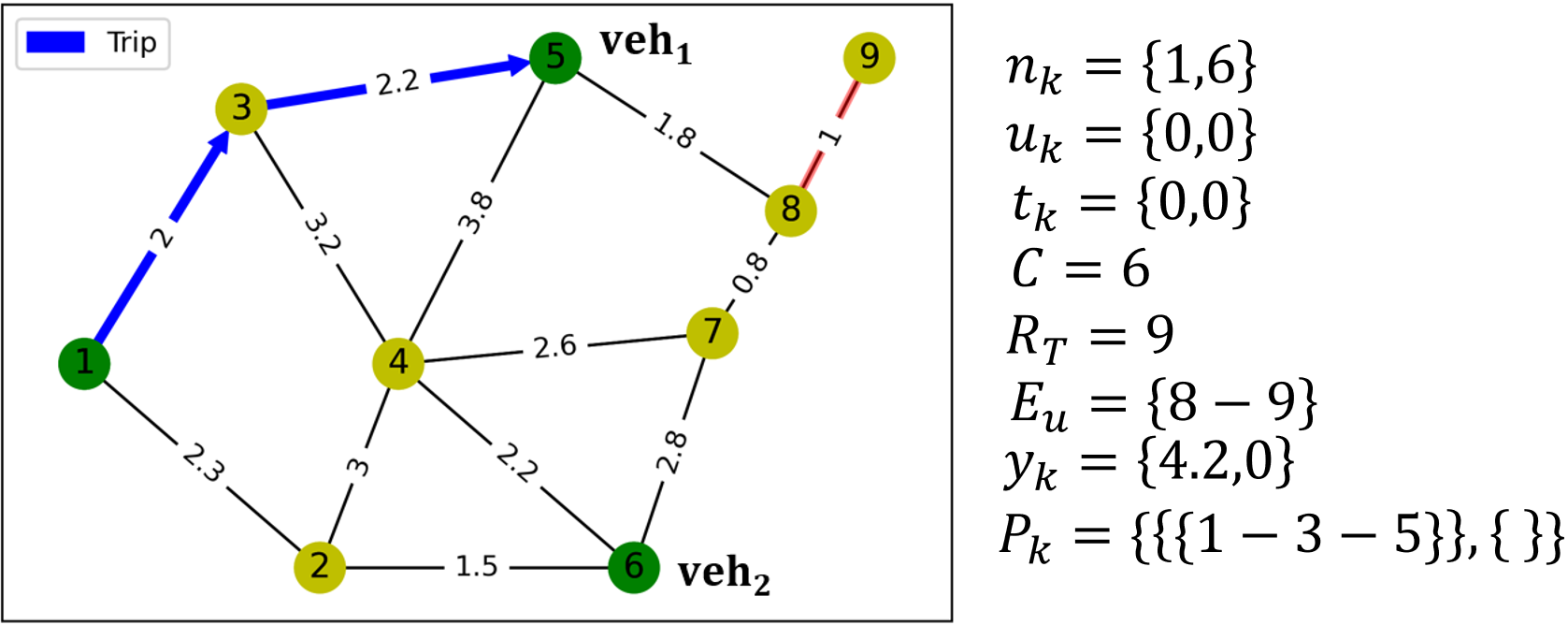}
    \caption{\textrm{Vehicle 1 Updated Location ($n_k = $ Vehicle Location, $u_k = $ Vehicle utilization, $t_k = $ Vehicle availability, $C = $ Vehicle capacity, $R_T = $ Recharge Time, $E_u$ = Required Edges), $P_k = $ Vehicle Path, $y_k = $ Vehicle last depot arrival)}}
    \label{sameexam2}
\end{figure}

Because neither vehicle can traverse the required edge in a single trip, the algorithm moves vehicle 1 from node 1 to node 5, which is closer to the required edge.
Figure \ref{sameexam2} shows the updated state after this move.

\begin{figure}[H]
    \centering
    \includegraphics[width=13.5cm]{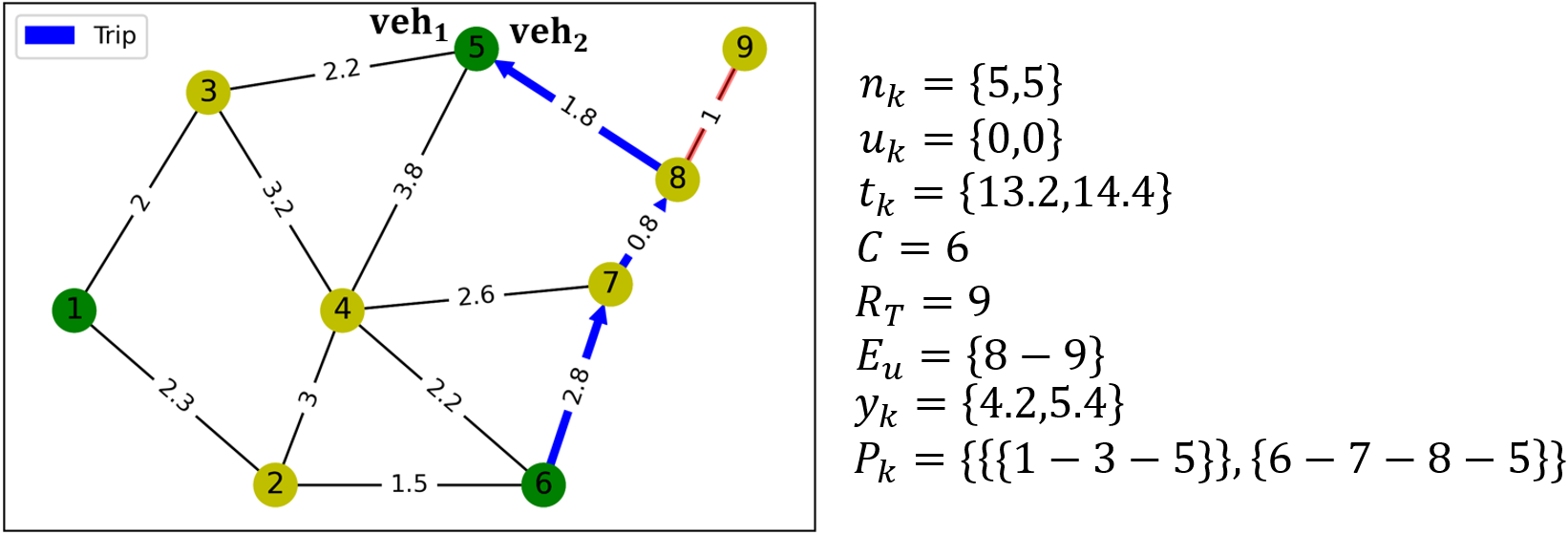}
    \caption{\textrm{Vehicle 2 Updated Location ($n_k = $ Vehicle Location, $u_k = $ Vehicle utilization, $t_k = $ Vehicle availability, $C = $ Vehicle capacity, $R_T = $ Recharge Time, $E_u$ = Required Edges), $P_k = $ Vehicle Path, $y_k = $ Vehicle last depot arrival)}}
    \label{sameexam3}
\end{figure}

Because Vehicle 1 needs to recharge after reaching depot node 5, Vehicle 2 is the first available vehicle in the next iteration. 
The algorithm moves Vehicle 2 from node 6 to node 5 in a single trip. 
Figure \ref{sameexam3} shows the updated state after this move.

\begin{figure}[H]
    \centering
    \includegraphics[width=13.5cm]{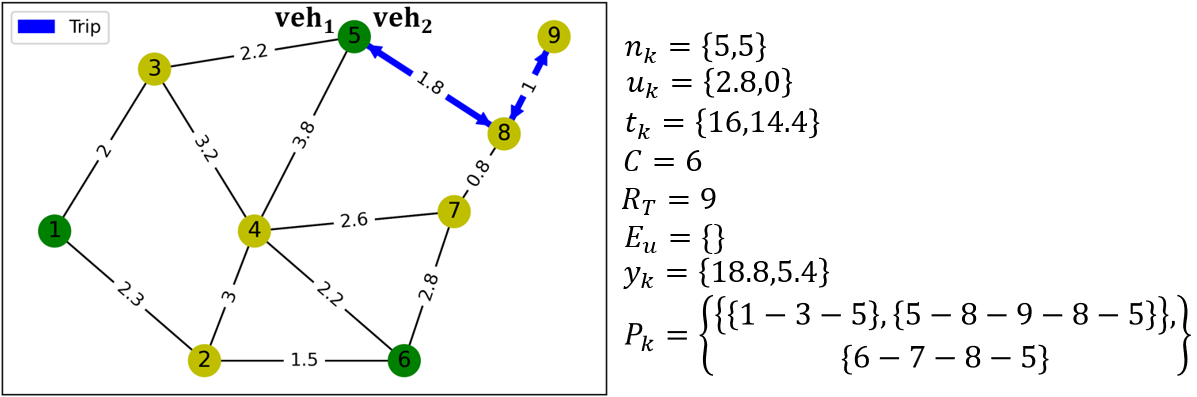}
    \caption{\textrm{Required Edge Traversal ($n_k = $ Vehicle Location, $u_k = $ Vehicle utilization, $t_k = $ Vehicle availability, $C = $ Vehicle capacity, $R_T = $ Recharge Time, $E_u$ = Required Edges), $P_k = $ Vehicle Path, $y_k = $ Vehicle last depot arrival)}}
    \label{sameexam4}
\end{figure}

In the next iteration, Vehicle 1 is the first available, and the algorithm extends its path to traverse the required edge in a trip that begins and ends at the depot at node 5. 
Figure \ref{sameexam4} shows this trip and the state after this trip.  
The maximum trip time equals $\max \{ 18.8, 5.4 \}$ = 18.8 time units.

This example highlights the fact that a vehicle may need to use multiple trips in order to traverse required edges. The following section discusses the results obtained from the multi-trip algorithm and existing heuristics and compares those solutions with the optimal solutions obtained by solving the MILP. 


\section{Results and Discussions}
\label{rd}
This section briefly describes some heuristics that have been used to solve the CARP; we used these as benchmarks for the multi-trip algorithm. 
This section provides details about the instances that we used in our experiments and describes some important steps followed to obtain optimal results using the Gurobi optimizer.
Finally, the section discusses the results from the selected heuristics,  our multi-trip algorithm, and the MILP, along with the analysis of the computational effort of the multi-trip algorithm.

\subsection{Existing CARP Heuristic Approaches}
\label{carpapp}
This section briefly reviews the following approaches for the CARP:  (i) the Path Scanning Algorithm, (ii) the Augment Merge Algorithm, and (iii) the Construct Strike Algorithm.

\subsubsection{Path Scanning (PS) Algorithm}
\label{ps}
The Path Scanning Algorithm \cite{Carp2} constructs vehicle routes that satisfy the constraints of the CARP by adding one edge at a time based on a myopic optimization criterion. Golden {\it et al.} \cite{Carp2} proposed five optimization criteria, and the criteria that produced the best routes is selected as the output to the algorithm. Although the myopic nature of the optimization criteria helps the algorithm to produce faster solutions, the quality of these solutions may be poor due to its greedy approach \cite{Carp3}. 

\subsubsection{Augment Merge (AM) Algorithm}
\label{am}
The Augment Merge algorithm \cite{Carp2} consists of two stages.  
The augment stage constructs routes by finding the shortest path from a depot to the nearest required arc and again from the required arc to the depot in such a way that each solution route traverses only one required arc and satisfies CARP constraints. 
The merge stage arranges the solution routes in descending order of the route cost.
Then, this stage combines the longest route with one of the smaller ones only if it satisfies the CARP constraints. 
Finally, this stage selects the combination with minimum total route cost as the output of  the algorithm. 
As the algorithm only considers two routes in the merging process, the resulting solution quality may improve compared to the path scanning algorithm with the increase in solution time, but solutions produced may not always be optimal as the merge step only tries to combine two routes \cite{Carp3}. \par

\subsubsection{Construct Strike (CS) Algorithm}
\label{cs}
The Construct Strike algorithm \cite{Carp5} iteratively performs three steps until all required edges are traversed. 
The first step uses the path scanning algorithm \cite{Carp2} to find a solution that satisfies the CARP constraints.  
The second step removes all of the edges in that solution from the graph. 
The third step determines whether the removed edges did not disconnect the graph. 
A graph is disconnected if there is no edge connecting the depot and the remaining graph. If the graph gets disconnected, artificial edges joining the depot and the remaining graph are added, and the graph with nodes and untraversed required edges is passed on to the first step. 
Although the construct strike algorithm avoids redundant traversal of required edges by removing all the edges in  the solution route from the graph once constructed, it also makes it challenging to construct solution routes as the iteration progresses due to the fewer number of edges in the graph, thus increasing the route cost. 
For this reason, the quality of the solutions may be poor. 
In some scenarios, the algorithm may fail to generate a solution that traverses all the required edges \cite{Carp3}. \par

Although these heuristics were proposed for the CARP problem with a single depot, they have been modified to work for multiple depots suitable for the MD-RPP-RV. The following subsection provides details on how the instances were generated for testing.
\subsection{Instance Generation Process}
\label{icrin}
We created instances from CARP benchmark instances \cite{Carp2} and from real-world road maps. 
The instance generation process randomly chose one-fifth of the nodes as depots and randomly chose one-third of the edges as required edges. 
Thus, the number of depots is less than the number of required edges. 
The instance generation process sets the number of vehicles as one-half of the number of required edges.  
This generated instances that were suitable for testing the proposed multi-trip algorithm and comparing them against the CARP heuristic-based approaches.
Sathyamurthy \cite{thesis} described the details of the instance generation process.

The following subsection provides information about the criteria and assumptions considered in the creation of different sets of instances used for obtaining results.

\subsubsection{Set A Instances}
\label{dear}
We created the Set A instances by converting 23 CARP instances with undirected graphs \cite{Carp2} to multi-depot capacitated RPP using the instance generation process. 
In these instances, the vehicle capacity was set to equal twice the maximum edge weight of the instance.
This was done so that a vehicle should be able to traverse the longest edge if it were chosen as one of the required edge.

\subsubsection{Set B Instances}
Set B has eight instances created from roadmaps of different regions with cold weather. 
We converted the roadmaps to undirected graphs and then to testing instances using the instance generation process. 
In these instances, the vehicle capacity is set to 31 minutes, the listed capacity of the DJI Mavic Pro 2 UAV \cite{dji}.

\subsubsection{Set C Instances}
Set C has eight instances created from the Set B instances by adding the effect of wind on the edge travel times, which requires relaxing assumption 8 (Subsection \ref{assump}). 
In these instances, each edge had two weights (travel times): one for one direction and a second for the opposite direction.
We determined the different weights by considering the impact of wind on a vehicle's travel time.
In one direction, the wind may increase the vehicle's speed (relative to the ground), which reduces the travel time; in the other direction, the wind may decrease the speed and increase the travel time.
Sathyamurthy \cite{thesis} described the details of these calculations. For set C instances, as only the graph type is changed from undirected to directed, we decided to keep the number of vehicles ($K$), number of required edges ($E_u$), and the number of depot nodes ($N_d$), the same as set B instances. For required edges ($E_u$), only one direction of the edge, i.e., either $(i,j)$ or $(j, i)$, is randomly chosen to be the required edge as edge weights in either direction are different.

 
\subsection{Optimal Solutions using Gurobi}
This section describes the procedure used to determine the number of trips ($F$) needed by Gurobi to produce optimal solutions and also modifications done to the instances provided as input to the Gurobi optimizer.

\subsubsection{Determining Number of Trips ($F$)}
\label{deter-trips}
From the MILP formulation (Section \ref{formulation}), in order to define the decision variables programmatically in Gurobi, fixed values for the number of vehicles ($K$) and the number of trips ($F$) are needed. But, $F$ cannot be known beforehand as each vehicle might take different number of trips to traverse required edges. Hence, $F$ needs to be an upper bound of the number of trips of $K$ vehicles. In order to find the upper bound, a simple iterative procedure is followed. \par

Initially, we set $F = 1$ and use Gurobi to find an optimal solution to the instance.  
 If a feasible solution exists, then the instance is solved.
If Gurobi reports that there are no feasible solutions, we increase $F$ by 1 and attempt to solve it again.  
This continues until Gurobi finds a feasible, optimal solution.

The number of variables and the Random Access Memory (RAM) required to solve the problem increase as $F$ increases.
By starting with $F = 1$ and increasing it only as needed, this approach avoids solving any unnecessarily large instances, which is more efficient.


 \subsubsection{Modified Gurobi Optimizer Input}
\begin{figure}[H]
    \centering
    \includegraphics[width=13.5cm]{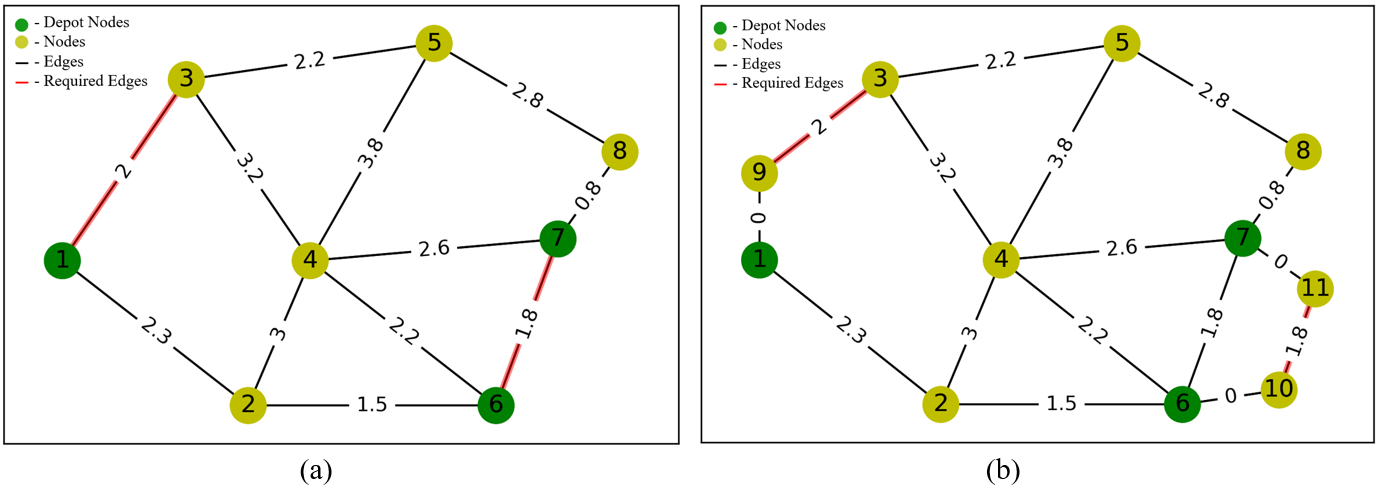}
    \caption{\textrm{(a) A MD-RPP-RV instance where required edges with one or both of its nodes as depot nodes, (b) Modified MD-RPP-RV instance with dummy nodes and edges.}}
    \label{instance-example}
\end{figure}

 Because of the subtour elimination constraint (Constraint 12), if a vehicle traverses a required edge in the subset $S$, then it also traverses edges that enter and leaves $S$. 
 This is true for all possible subsets formed by the node set $N/N_d$. 
A problem occurs  if a required edge has one or both of its nodes as depot nodes (Figure \ref{instance-example}{\color{blue}a}). 
In this scenario, there are no subsets possible as subsets do not include depot nodes ($N_d$). 
To avoid this scenario, we added dummy nodes and edges with zero edge weights to modify any instance with such a required edge. 
For example, in Figure \ref{instance-example}{\color{blue}b}, the required edge $(1,3)$ has node 1 as the depot node, and hence, a dummy node 9 and dummy edge $(1,9)$ is inserted with an edge weight of 0, and the required edge $(1,3)$ changes to $(9,3)$ with the same edge weight. Note that the insertion of dummy nodes and edges has no effect on the quality of routes produced. A similar example is shown in Figure \ref{instance-example}{\color{blue}a} with the required edge $(6,7)$ where both of its nodes are depot nodes.

\subsection{Results}
This section presents the results from our tests of the proposed multi-trip algorithm, existing heuristic-based approaches, and solving the MILP  using  Gurobi  on the instances in Sets A, B, and C. 
These tests were run on an AMD EPYC 7763 64-Core Processor with 128 physical cores, 128 logical processors, and 8 CPU cores, using up to 32 threads with 8 GB memory per CPU core.

\subsubsection{Benchmark Instances Results}
 Table \ref{abrre} provides a list of abbreviations used, and Table \ref{benchmark-instance} and \ref{benchmark-results} describe the benchmark instances and provide the results obtained, respectively.

\begin{table}[H]\
\centering
\renewcommand{\arraystretch}{1}
\caption{\textrm{Abbreviations}}
\label{abrre}
\textrm{
\resizebox{7.3cm}{!}{
\begin{tabular}{cc}
\hline
\multicolumn{1}{c}{Abbreviation} & \multicolumn{1}{c}{Expansion}   \\ \hline
PS                                 & Path Scanning Algorithm          \\
CS                                 & Construct Strike Algorithm       \\
AM                                 & Augment Merge Algorithm          \\
MT                               & Multi-trip Algorithm \\
$tu$                                 & Time units \\
M                                 & Maximum Trip Time \\
ET                                 & Execution Time \\\hline
\end{tabular}}}
\end{table}

\begin{table}[H]
\centering
\renewcommand{\arraystretch}{1}
\caption{\textrm{Set A Instances Information ($N = $ Number of Nodes, $E = $ Number of Edges, $E_{u} = $ Number of Required Edges, $K = $ Number of Vehicles, $C = $ Vehicle Capacity, $N_d = $ Number of Depots 
)}}
\label{benchmark-instance}
\textrm{
\resizebox{5.5cm}{!}{
\begin{tabular}{crrrrrr}
\hline
\multicolumn{1}{c}{\begin{tabular}[c]{@{}c@{}}Instance \\   Name\end{tabular}} & \multicolumn{1}{c}{$N$} & \multicolumn{1}{c}{$E$} & \multicolumn{1}{c}{$E_u$} & \multicolumn{1}{c}{$C$} & \multicolumn{1}{c}{$K$} & \multicolumn{1}{c}{$N_d$} \\ \hline
A.1                                                                          & 8                      & 11                     & 3                       & 18                     & 1                      & 2                       \\
A.2                                                                          & 11                     & 19                     & 5                       & 40                     & 2                      & 3                       \\
A.3                                                                          & 7                      & 21                     & 8                       & 16                     & 4                      & 2                       \\
A.4                                                                          & 7                      & 21                     & 8                       & 12                     & 4                      & 2                       \\
A.5                                                                          & 12                     & 22                     & 7                       & 40                     & 3                      & 3                       \\
A.6                                                                          & 11                     & 22                     & 8                       & 18                     & 4                      & 3                       \\
A.7                                                                          & 12                     & 22                     & 8                       & 40                     & 4                      & 3                       \\
A.8                                                                          & 12                     & 22                     & 7                       & 44                     & 3                      & 3                       \\
A.9                                                                          & 12                     & 22                     & 8                       & 40                     & 4                      & 3                       \\
A.10                                                                         & 13                     & 23                     & 7                       & 60                     & 3                      & 3                       \\
A.11                                                                         & 12                     & 25                     & 9                       & 38                     & 4                      & 3                       \\
A.12                                                                         & 12                     & 26                     & 9                       & 40                     & 4                      & 3                       \\
A.13                                                                         & 13                     & 26                     & 7                       & 44                     & 3                      & 3                       \\
A.14                                                                         & 10                     & 28                     & 9                       & 198                    & 4                      & 3                       \\
A.15                                                                         & 8                      & 28                     & 10                      & 16                     & 5                      & 2                       \\
A.16                                                                         & 8                      & 28                     & 10                      & 14                     & 5                      & 2                       \\
A.17                                                                         & 11                     & 33                     & 11                      & 18                     & 5                      & 3                       \\
A.18                                                                         & 9                      & 36                     & 13                      & 16                     & 6                      & 2                       \\
A.19                                                                         & 11                     & 44                     & 14                      & 18                     & 7                      & 3                       \\
A.20                                                                         & 22                     & 45                     & 16                      & 38                     & 8                      & 5                       \\
A.21                                                                         & 27                     & 46                     & 13                      & 18                     & 6                      & 6                       \\
A.22                                                                         & 27                     & 51                     & 16                      & 18                     & 8                      & 6                       \\
A.23                                                                         & 11                     & 55                     & 18                      & 16                     & 9                      & 3                       \\ 
\hline
\end{tabular}}}
\end{table}

\begin{table}[H]
\centering
\renewcommand{\arraystretch}{1}
\caption{\textrm{Benchmark Instances Results}}
\label{benchmark-results}
\textrm{
\resizebox{11.5cm}{!}{
\begin{tabular}{rrrrrrrrrrrr}
\hline
\multicolumn{1}{c}{\multirow{2}{*}{\begin{tabular}[c]{@{}c@{}}Instance  \\  Name\end{tabular}}} & \multicolumn{2}{c}{PS results}                 & \multicolumn{2}{c}{CS results}                 & \multicolumn{2}{c}{AM results}                 & \multicolumn{2}{c}{MT Results}                 & \multicolumn{2}{l}{MILP Results}               & \multicolumn{1}{c}{\multirow{2}{*}{Gap \%}} \\
\multicolumn{1}{c}{}                                                                            & \multicolumn{1}{c}{ET} & \multicolumn{1}{c}{M} & \multicolumn{1}{c}{ET} & \multicolumn{1}{c}{M} & \multicolumn{1}{c}{ET} & \multicolumn{1}{c}{M} & \multicolumn{1}{c}{ET} & \multicolumn{1}{c}{M} & \multicolumn{1}{c}{ET} & \multicolumn{1}{c}{M} & \multicolumn{1}{c}{}                        \\ \hline
A.1                                                                                             & 0.0                    & 214                   & --                     & \textbf{--}           & 0.0                    & 236                   & 0.0                    & 64                    & 0.1                    & 64                    & 0.0                                         \\
A.2                                                                                             & 0.0                    & 264                   & 0                      & 264                   & 0.0                    & 264                   & 0.0                    & 244                   & 117.1                  & 148                   & 64.9                                        \\
A.3                                                                                             & 0.0                    & 113                   & --                     & \textbf{--}           & 0.0                    & 113                   & 0.0                    & 16                    & 6.7                    & 15                    & 6.7                                         \\
A.4                                                                                             & 0.0                    & 100                   & 0                      & 197                   & 0.0                    & 102                   & 0.0                    & 32                    & 166.4                  & 8                     & 300.0                                       \\
A.5                                                                                             & 0.0                    & 233                   & 0                      & 273                   & 0.0                    & 297                   & 0.0                    & 136                   & 3255.8                 & 126                   & 7.9                                         \\
A.6                                                                                             & 0.0                    & 115                   & --                     & \textbf{--}           & 0.0                    & 115                   & 0.0                    & 138                   & 76.6                   & 16                    & 281.3                                       \\
A.7                                                                                             & 0.0                    & 148                   & --                     & \textbf{--}           & 0.0                    & 167                   & 0.0                    & 163                   & 339.6                  & 34                    & 305.9                                       \\
A.8                                                                                             & 0.0                    & 257                   & --                     & \textbf{--}           & 0.0                    & 314                   & 0.0                    & 149                   & 75.9                   & 42                    & 288.1                                       \\
A.9                                                                                             & 0.0                    & 263                   & --                     & \textbf{--}           & 0.0                    & 263                   & 0.0                    & 61                    & 2788.1                 & 134                   & 11.2                                        \\
A.10                                                                                            & --                     & --                    & --                     & \textbf{--}           & --                     & --                    & 0.0                    & 361                   & 211.3                  & 59                    & 511.9                                       \\
A.11                                                                                            & 0.0                    & 244                   & --                     & \textbf{--}           & 0.0                    & 285                   & 0.0                    & 121                   & 65486.5                      & 120                     & 0.8                                           \\
A.12                                                                                            & 0.0                    & 245                   & --                     & \textbf{--}           & 0.0                    & 245                   & 0.0                    & 148                   & 20170.9                      & 110                     & 34.5                                           \\
A.13                                                                                            & --                     & --                    & --                     & \textbf{--}           & --                     & --                    & 0.0                    & 166                   & 22853.3                      & 141                     & 17.7                                           \\
A.14                                                                                            & 0.0                    & 255                   & 0                      & 350                   & 0.0                    & 337                   & 0.0                    & 63                    & 434.4                  & 48                    & 31.3                                        \\
A.15                                                                                            & 0.0                    & 110                   & --                     & \textbf{--}           & 0.0                    & 114                   & 0.0                    & 16                    & 273.2                  & 13                    & 23.1                                        \\
A.16                                                                                            & 0.0                    & 104                   & --                     & \textbf{--}           & 0.0                    & 110                   & 0.0                    & 13                    & 547.1                  & 10                    & 30.0                                        \\
A.17                                                                                            & 0.0                    & 201                   & --                     & \textbf{--}           & 0.0                    & 209                   & 0.0                    & 52                    & *                     & *                    & --                                          \\
A.18                                                                                            & 0.0                    & 203                   & --                     & \textbf{--}           & 0.0                    & 226                   & 0.0                    & 16                    & 55548.1                      & 13                     & 23.1                                           \\
A.19                                                                                            & 0.0                    & 106                   & 0                      & 113                   & 0.0                    & 118                   & 0.1                    & 48                    & *                    & *                   & --                                          \\
A.20                                                                                            & 0.0                    & 146                   & --                     & --                    & 0.0                    & 154                   & 0.1                    & 31                    & *                    & *                   & --                                          \\
A.22                                                                                            & 0.0                    & 208                   & --                     & --                    & 0.0                    & 219                   & 0.0                    & 66                    & *                    & *                   & --                                          \\
A.22                                                                                            & 0.0                    & 436                   & --                     & --                    & 0.0                    & 436                   & 0.1                    & 103                   & *                    & *                   & --                                          \\
A.23                                                                                            & 0.0                    & 106                   & --                     & --                    & 0.0                    & 115                   & 0.1                    & 44                    & *                    & *                   & --                                          \\ \cline{1-9} \hline
\end{tabular}}}
\end{table}

In Table \ref{benchmark-results}, the algorithms that did not completely solve an instance are shown with a single dash ($-$). 
For some small instances in Set A, the execution time of some algorithms was less than 50 milliseconds; these appear as $0.0$ in Table \ref{benchmark-results} because the execution times were rounded to one decimal point. For some of the set A instances, the Gurobi optimizer could not produce optimal solutions using the MILP formulation as it ran out of allocated memory after running for more than 30 hours. For these instances, the results are represented as * in the table. 

To measure the quality of the solutions the multi-trip algorithm constructed relative to the quality of the optimal solutions found by solving the MILP using Gurobi, we calculated the relative difference as follows.
Let $M_{MT}$ be the maximum trip time of a solution from the multi-trip algorithm.
Let $M_{MILP}$ be the maximum trip time of an optimal solution of the MILP.
The metric Gap is the percentage relative difference between these values: 
\begin{equation}
    \text{Gap} = \frac{M_{MT} - M_{MILP}}{M_{MILP}} \times 100 
\end{equation}

The average gap considering only instances we obtained optimal solutions for set A instances was 114 \%. 
Although the deviation of solution quality of the MT algorithm with respect to the optimum is large, the trade-off comes with significant execution time taken to obtain the optimal solution. For example, for set A instance A.13 in Table \ref{benchmark-results}, the MT algorithm solved the instance within a second, but the Gurobi optimizer took more than 6 hours to obtain the optimal solution with a gap of 17.7 \%, which emphasizes the importance of MT algorithms' ability to produce quick routes. The MT algorithm also outperformed all heuristic-based approaches from the literature by obtaining better solutions in all set A instances except for instances A.6 and A.7. The following subsection provides the real-world instances results.
\subsubsection{Real-world Instances Results}
Tables \ref{rw-instances} and \ref{rw-results} describe the real-world instances and  the results obtained on those instances. 
Because these instances had many more nodes and edges than the benchmark instances, we were unable to obtain optimal solutions to the MILP  using Gurobi due to the large compute and memory requirements. 
\begin{table}[H]
\centering
\renewcommand{\arraystretch}{1}
\caption{\textrm{Set B and C Instances Information ($N = $ Number of Nodes, $E = $ Number of Edges, $E_{u} = $ Number of Required Edges, $K = $ Number of Vehicles, $C = $ Vehicle Capacity, $N_d = $ Number of Depots 
)}}
\label{rw-instances}
\textrm{
\resizebox{6.5cm}{!}{
\begin{tabular}{crrrrrr}
\hline
\multicolumn{1}{c}{\begin{tabular}[c]{@{}c@{}}Instance \\   Name\end{tabular}} & \multicolumn{1}{c}{$N$} & \multicolumn{1}{c}{$E$} & \multicolumn{1}{c}{$E_u$} & \multicolumn{1}{c}{$C$} & \multicolumn{1}{c}{$K$} & \multicolumn{1}{c}{$N_d$} \\ \hline
B.1                                                                          & 75                     & 130                    & 39                      & 31                     & 19                     & 16                      \\
B.2                                                                          & 133                    & 214                    & 28                      & 31                     & 14                     & 27                      \\
B.3                                                                          & 222                    & 344                    & 115                     & 31                     & 57                     & 45                      \\
B.4                                                                          & 192                    & 353                    & 113                     & 31                     & 56                     & 39                      \\
B.5                                                                          & 290                    & 423                    & 44                      & 31                     & 22                     & 59                      \\
B.6                                                                          & 288                    & 494                    & 163                     & 31                     & 81                     & 58                      \\
B.7                                                                          & 374                    & 622                    & 85                      & 31                     & 42                     & 75                      \\
B.8                                                                          & 461                    & 879                    & 286                     & 31                     & 143                    & 93                      \\ 
C.1                                                                          & 75                     & 260                    & 39                      & 31                     & 19                     & 16                      \\
C.2                                                                          & 133                    & 428                    & 28                      & 31                     & 14                     & 27                      \\
C.3                                                                          & 222                    & 688                    & 115                     & 31                     & 57                     & 45                      \\
C.4                                                                          & 192                    & 706                    & 113                     & 31                     & 56                     & 39                      \\
C.5                                                                          & 290                    & 846                    & 26                      & 31                     & 13                     & 59                      \\
C.6                                                                          & 288                    & 988                    & 163                     & 31                     & 81                     & 58                      \\
C.7                                                                          & 374                    & 1244                   & 81                      & 31                     & 40                     & 75                      \\
C.8                                                                          & 461                    & 1758                   & 286                     & 31                     & 143                    & 93                     \\ 
\hline
\end{tabular}}}
\end{table}

\begin{table}[H]
\centering
\renewcommand{\arraystretch}{1}
\caption{\textrm{Set B and C Instances Results (B.1-B.8 $=$ Set B Instances,  C.1-C.8 $=$ Set C Instances, ET $=$ Execution Time in seconds, M $=$ Maximum Trip Time in minutes)}}
\label{rw-results}
\textrm{
\resizebox{9.5cm}{!}{
\begin{tabular}{rrrrrrrrr}
\hline
                                                           & \multicolumn{2}{c}{PS results} & \multicolumn{2}{c}{CS results} & \multicolumn{2}{c}{AM results} & \multicolumn{2}{c}{MT Results} \\
\begin{tabular}[c]{@{}c@{}}Instance  \\  Name\end{tabular} & ET             & M             & ET        & M                  & ET             & M             & ET             & M             \\
\hline
B.1                                                        & 2              & 216           & --        & \textbf{--}        & 2              & 203           & 3.2            & 28.7          \\
B.2                                                        & --             & --            & --        & --                 & --             & --            & 0.4            & 356.7         \\
B.3                                                        & 60             & 194           & --        & \textbf{--}        & 55             & 124           & 292.3          & 30.8          \\
B.4                                                        & --             & --            & --        & --                 & --             & --            & 29.0           & 127.2         \\
B.5                                                        & --             & --            & --        & --                 & --             & --            & 1.2            & 563.0         \\
B.6                                                        & 107            & 223           & --        & \textbf{--}        & 103            & 223           & 313.4          & 130.3         \\
B.7                                                        & --             & --            & --        & \textbf{--}        & --             & --            & 4.3            & 593.6         \\
B.8                                                        & 158            & 227           & --        & \textbf{--}        & 174            & 227           & 224.3          & 122.6         \\
C.1                                                        & 8              & 214           & --        & \textbf{--}        & 4              & 123           & 5.6            & 25.2          \\
C.2                                                        & --             & --            & --        & \textbf{--}        & --             & --            & 0.8            & 578.2         \\
C.3                                                        & --             & --            & --        & \textbf{--}        & 164            & 207           & 492.3          & 28.6          \\
C.4                                                        & --             & --            & --        & \textbf{--}        & --             & --            & 41.8           & 135.8         \\
C.5                                                        & 0              & 241           & --        & \textbf{--}        & --             & --            & 0.7            & 808.1         \\
C.6                                                        & 360            & 246           & --        & --                 & 329            & 246           & 447.3          & 30.7          \\
C.7                                                        & 1              & 656           & --        & \textbf{--}        & --             & --            & 3.4            & 449.3         \\
C.8                                                        & 398            & 228           & --        & \textbf{--}        & 588            & 228           & 319.7          & 122.2     \\   \hline
\end{tabular}}}
\end{table}

In Table \ref{rw-results}, the algorithms that did not completely solve an instance are shown with a single dash ($-$).
The proposed multi-trip algorithm solved all of the real-world instances and constructed better solutions  than the CARP heuristic-based approaches. 
 The CARP heuristic-based approaches combined could solve only 10  of 16 real-world instances. 
On average, the multi-trip algorithm solved the real-world instances (which had 75 to 461 nodes and 130 to 1758 edges) within 3 minutes (136.2 seconds). 

\subsection{Analysis of Computational Effort}
\label{dis}
The analysis of the theoretical computational complexity helps better understand the algorithm's efficiency and scalability. In this regard, the multi-trip algorithm's theoretical upper and lower bounds play a crucial role in assessing its performance and suitability for various real-world applications.

The upper bound ($O$) provides a worst-case estimate of the algorithm's runtime as a function of inputs (Table \ref{tb1}). This information is essential in determining the algorithm's scalability and identifying scenarios where the algorithm may fail to meet the desired performance requirements. On the other hand, the lower bound ($\Omega$) provides a theoretical guarantee on the minimum number of operations required to solve the problem by analyzing the best-case inputs. This information is useful in assessing the algorithm's inherent complexity and identifying opportunities for optimization. \\

\textbf{Claim 1} \textit{The theoretical upper bound for MT algorithm is $O(|E|^2|N|^2)$},\\
where $|N|$ and $|E|$ are the number of nodes and edges of the MD-RPP-RV instance.\\ \par
\textbf{Claim 2} \textit{The theoretical lower bound for MT algorithm is $\Omega(K)$},\\
where $K$ is the number of vehicles. \\

The proof of claims 1 and 2 are provided in Appendix.


\section{Conclusions}
\label{conc}
This paper defined and formulated the MD-RPP-RV problem as a MILP  and presented a new heuristic approach, the multi-trip algorithm. 
This algorithm was developed by studying and gaining insights from the shortcomings of heuristic-based approaches that were developed to solve the CARP.

In order to test the multi-trip algorithm and compare it with the heuristic-based approaches, we created instances from the literature and from real-world roadmaps. In addition to this, we created instances that modeled the effects of wind on vehicle travel times, which is relevant to the winter weather scenarios that motivated this research.

We obtained optimal solutions for the smaller instances by solving the MILP using Gurobi, but we could not find optimal solutions for the larger, real-world instances  due to memory and computing limitations.
The results showed that the proposed multi-trip algorithm produced better-quality solutions than the heuristic-based approaches for all instances. The average gap between the quality of the solutions from the  multi-trip algorithm and the quality of the optimal solutions was 114\%. 
The multi-trip algorithm solved the larger real-world instances on an average within 3 minutes.   


In terms of future directions, there are many assumptions that are implicitly considered, such as no vehicle failures during operation, 
that might be unrealistic in some situations.
One possible future direction would be to consider the problem with multiple vehicle failures and to develop approaches that can construct robust solutions.
Another possible future direction would be to model temperature effects on vehicle batteries and integrate them with the multi-trip algorithm to produce realistic results. 
The facility location problem (determining the best places for depots to minimize the total time required) is also interesting.

\section*{Acknowledgement}
This research is funded by the Maryland Industrial Partnerships (MIPS) program at the University of Maryland and by SpringGem.
The work of the first author was supported in part by a generous gift from Dr.\ Alex Mehr (Ph.D. ’03) through the Design Decision Support Laboratory Research and Education Fund and by Army Cooperative Agreement W911NF2120076. We would like to acknowledge Dr.\ Menglin Jin from SpringGem for the help and support. We would also like to acknowledge Dr.\ Michael Otte for providing information and feedback on the research.  The authors acknowledge the University of Maryland High-Performance Computing resources (\href{http://hpcc.umd.edu}{{\color{blue}Zaratan}}) made available for conducting the research reported in this article.






\appendix
\section*{}
\scalebox{1.3}{\textbf{\appendixname}}
\label{appendix}
\\ \\
\textbf{\textit{Proof of Claim 1:}} \par
In order to prove the theoretical upper bound of the proposed multi-trip algorithm, the worst-case instance of the MD-RPP-RV is considered, where all vehicles need to make multi-trip routes to traverse the required edges. One such instance is described and solved using the proposed multi-trip algorithm in Subsection \ref{demo}. \par
In the worst-case scenario, the multi-trip algorithm, as described in the pseudocode (Section \ref{v1}), will remain in the while loop (line 1) till all required edges ($E_u$) are traversed, with the number of operations needed as $O(|E_u|)$. Inside the loop in line 2, it takes at most takes $K$ steps to find the quickest available vehicle, i.e., $O( |E_u| \times (K))$. For determining the closest feasible edge from the vehicle location $n_k$ in line 3, all $|E_u|$ required edges must be considered to determine the closest feasible edge. As the multi-trip algorithm uses the Dijkstra algorithm to determine these distances, which has a complexity of $O(|N|^2)$, the overall complexity becomes $O(|E_u| \times (K + |E_u| \times |N|^2))$. \par

The algorithm then enters the else statement in line 14, as no single trip route is possible. To determine the closest required edge from the set $E_u$ in line 15, it takes $|N|^2|E_u|$ again steps using Dijkstra algorithm, $O(|E_u| \times (K + 2 \times |E_u| \times |N|^2))$. In line 16, determining the closest feasible depot is similar to determining the closest feasible edge. So, every depot node needs to be considered to determine the closest depot node from $n_k$, which has a complexity of $O(|N_d| \times |N|^2)$ bringing the total complexity of the algorithm to $O(|E_u| \times (K + 2 \times |E_u| \times |N|^2 + |N_d| \times |N|^2))$. Finally, in line 18, determining the shortest path to $d^{*}$ takes $|N|^2$, $O(|E_u| \times (K + 2 \times |E_u| \times |N|^2 + |N_d| \times |N|^2 + |N|^2))$. The algorithm exists while loop but needs to complete the routes of vehicles (line 26) by finding the closest depot (line 28), and the total complexity of the algorithm becomes $O(|E_u| \times (K + 2 \times |E_u| \times |N|^2 + |N_d| \times |N|^2 + |N|^2) + K \times |N_d| \times |N|^2)$. \par

But, based on how the instances were generated (Section \ref{icrin}) for testing, we know $|E_u| = \frac{|E|}{3}$, $K = \frac{|E_U|}{2} = \frac{|E|}{6}$ and $|N_d| = \frac{|N|}{5}$. Hence, we can represent the upper bound of the algorithm compactly in terms of the number of nodes and edges. 
\begin{alignat*}{1}
    &O(|E_u| \times (K + 2 \times |E_u| \times |N|^2 + |N_d| \times |N|^2 + |N|^2) + K \times |N_d| \times |N|^2)\\
    &O(\frac{|E|}{3} \times (\frac{|E|}{6} + 2 \times \frac{|E|}{3} \times |N|^2 + \frac{|N|}{5} \times |N|^2 + |N|^2) + \frac{|E|}{6} \times \frac{|N|}{5} \times |N|^2)\\
    &O(\frac{|E|^2}{18} + 2 \times \frac{|E|^2}{9} \times |N|^2 + \frac{|N|^3}{5} \times \frac{|E|}{3} + |N|^2 \times \frac{|E|}{3} + \frac{|E|}{6} \times \frac{|N|^3}{5})\\
\end{alignat*}
Removing the lower-order terms and constants, the upper bound simplifies to:
\begin{alignat*}{1}
    &O(|E|^2 \times |N|^2 + |N|^3 \times |E|)\\
\end{alignat*}
As most of the instances are a multi-graph where multiple edges between nodes exist, $|E| \geq |N|$. Hence, $|E|^2 \times |N|^2 \geq |N|^3 \times |E|$. Therefore, the upper bound of the multi-trip algorithm becomes:
\begin{alignat*}{1}
    &O(|E|^2 |N|^2)
\end{alignat*}
\\
\textbf{\textit{Proof of Claim 2:}} \par
To prove the theoretical lower bound, the best-case scenario is considered, which is there are no required edges to traverse. In this case, the multi-trip algorithm enters directly into line 26 and loops through all vehicles to check for incomplete routes. This looping takes $K$ steps, and hence, the lower bound of the multi-trip algorithm is given by:
\begin{alignat*}{1}
    &\Omega(K)
\end{alignat*}

\end{document}